\documentclass[preprint,showpacs,showkeys,preprintnumbers,amsmath,amssymb]{revtex4}

\usepackage{graphicx}
\usepackage{dcolumn}
\usepackage{bm}

\newcommand {\nn}    {\nonumber}

\begin{document}

\title{Fermions in Self-dual Vortex Background on a String-like Defect}

\author{Yu-Xiao Liu}
\thanks{Corresponding author} \email{liuyx@lzu.edu.cn}
\author{Li Zhao}
\email{zhl03@lzu.cn}
\author{Xin-Hui Zhang}
\email{zhangxingh03@lzu.cn}
\author{Yi-Shi Duan}
\email{ysduan@lzu.edu.cn}

\affiliation{Institute of Theoretical Physics, Lanzhou University,
Lanzhou 730000, P. R. China}

\begin{abstract}
By using the self-dual vortex background on extra two-dimensional
Riemann surfaces in 5+1 dimensions, the localization  mechanism of
bulk fermions on a string-like defect with the exponentially
decreasing warp-factor is obtained. We give the conditions under
which localized spin 1/2 and 3/2 fermions can be obtained.
\end{abstract}

 %  \\

%11.10.Kk   Field theories in dimensions other than four
%04.50.+h   Gravity in more than four dimensions, Kaluza-Klein
%           theory, unified field theories; alternative theories of gravity
%           (see also 11.25.Mj Compactification and four-dimensional models)

\date{\today}

\maketitle

\section{Introduction}

It is now widely believed that extra dimensions play an important
role in constructing a unified theory of all interactions and
provides us with a new solution to hierarchy problem
\cite{hierarchy,RS}. The possible existence of such dimensions got
strong motivation from theories that try to incorporate gravity
and  gauge interactions in a unique scheme, in a reliable manner.
The idea dates back to the 1920's,  to the works of Kaluza and
Klein \cite{KaluzaKlein} who tried to unify electromagnetism with
Einstein gravity by assuming that the photon originates from the
fifth component of the metric.

Recently, co-dimension two models in six dimensions have been a
topic of increasing interest
\cite{6Dmodels,6dmodel,GherghettaPRL2000,OdaPLB2000113}. Apart
from model construction, the question of solving the cosmological
constant problem has been the primary issue addressed in several
articles \cite{cosmoconstant6d}. Other aspects such as cosmology,
brane gravity {\cite{codim2gen}, fermion families and chirality
{\cite{ErdemEPJC2002} etc. have been discussed by numerous
authors. A list of some recent articles on codimension two models
is provided in Ref. \cite{codim2recent}.

In the brane world scenario, our universe is regarded as a 3-brane
embedded in a higher-dimensional space-time  with non-factorizable
warped geometry. It is a priori assumed that all the matter fields
are constrained to live on the three brane, whereas gravity is
free to propagate in the extra dimension. Then a key ingredient
for realizing the brane world idea is localization of various bulk
fields on a brane by a natural mechanism. In other words, in this
scenario various fields we observe in our universe are nothing but
the zero modes of the corresponding bulk fields which are trapped
on our brane by some ingenious mechanism.

This localization mechanism has been recently investigated within
the framework of a local field theory. Ever since Goldberger and
Wise \cite{goldwise} added a bulk scalar field to fix the location
of the branes in five dimensions, investigations with bulk fields
became an active area of research. It has been shown that the
graviton \cite{RS} and the massless scalar field
\cite{BajcPLB2000} have normalizable zero modes on branes of
different types, that the Abelian vector fields are not localized
in the Randall-Sundrum (RS) model in five dimensions but can be
localized in some higher-dimensional generalizations of it
\cite{OdaPLB2000113}. In contrast, in
\cite{BajcPLB2000,NonLocalizedFermion} it was shown that fermions
do not have normalizable zero modes in five dimensions, while in
\cite{OdaPLB2000113} the same result was derived for a
compactification on a string \cite{GherghettaPRL2000} in six
dimensions. Subsequently, Randjbar-Daemi {\it et al} studied
localization of bulk fermions on a brane with inclusion of scalar
backgrounds \cite{RandjbarPLB2000} and minimal gauged supergravity
\cite{Parameswaran0608074} in higher dimensions and gave the
conditions under which localized chiral fermions can be obtained.

Since spin half fields can not be localized on the brane
\cite{RS,OdaPLB2000113} in five or six dimensions by gravitational
interaction only, it becomes necessary to introduce additional
non-gravitational interactions to get spinor fields confined to
the brane or string-like defect. Fermionic zero modes in the
absence of gravity and in four dimensions in vortex background
were studied in \cite{Jackiw} and  extended to the case of six
dimensional space-times in gravity, gauge and vortex backgrounds
in \cite{liuyx}. The aim of the present article is to study
localization of bulk fermions on a string-like defect with
codimension 2 in self-dual vortex background.  In this article, we
first review the solutions  to Einstein's equations with a warp
factor in a 6-dimensional space-time, which has been studied by
many groups \cite{6Dmodels,GherghettaPRL2000,Vilenkin,EinsteinEq}.
Then, we shall prove that spin 1/2 and 3/2 fields can be localized
on a defect with the exponentially decreasing warp factor if the
self-dual vortex and gravitational backgrounds are considered.

\section{Self-dual vortex on a two-dimensional curved space}
\label{SectionVortex}

%We consider a (5+1)-dimensional space-time $M^{4}\times{K^2}$ with
%$M^{4}$ represents our four-dimensional space-time and $K^2$
%represents a two-dimensional extra Riemann surface. The metric
%$G_{MN}$ of the manifold $M^{4}\times{K^2}$ is determined by

This paper is focused on braneworld models with codimension
greater than one. In particular, we shall be exclusively concerned
with bulk spacetimes in six dimensions generically represented by
the line element
\begin{eqnarray}
ds^2 &=& g_{MN}dx^M dx^N \nn \\
     &=& g_{\mu \nu}(x,y)dx^\mu{d}x^\nu
      +  \gamma_{ij}(y)dy^{i}dy^{j}, \label{metric}
\end{eqnarray}
where $M, N$ denote 6-dimensional space-time indices,
$\mu,\nu=0,1,2,3$ and $i,j=1,2$ for our 4-dimensional space-time
and the two dimensional extra space $K^2$, respectively,
$\gamma_{ij}$ is the metric on $K^2$.

To generate the vortex solution, we introduce the generalized
Abelian Higgs Lagrangian
\begin{equation}
\mathcal{L}_{AH}=\sqrt{-g} \left
(-\frac{1}{4}F_{MN}F^{MN}+(D^{M}\phi)^{\dag}(D_{M}\phi)-\frac{\lambda}{2}(\|\phi\|^{2}-v^{2})^{2}
\right ),\label{lagrangianAH}
\end{equation}
where $g=\det(g_{MN})$,
$F_{MN}=\partial_{M}A_{N}-\partial_{N}A_{M}$, $\phi=\phi(y^k)$ is
a complex scalar field on extra dimensions,
$\|\phi\|=(\phi\phi^{\ast})^{\frac{1}{2}}$, $A_{M}$ is a U(1)
gauge field, $D_{M}=\partial_{M}-ieA_{M}$ is gauge-covariant. In
Eq. (\ref{lagrangianAH}), $v$ is the vacuum expectation value of
the Higgs field determining the masses of the Higgs and of the
gauge boson
\begin{equation}
m_H=\sqrt{2\lambda}v,~~~ m_V=ev.
\end{equation}
The Abrikosov-Nielsen-Olesen vortex solution on $K^2$ could be
generated from the Higgs field $\phi$. In the generalized Abelian
Higgs model, if the system admits a Bogomol'nyi limit
\cite{BogoSJNP1976}, one can arrive at the first-order Bogomol'nyi
self-duality equations in a curved space \cite{KimJMP2002}:
\begin{eqnarray}
&&B=\mp e(\|\phi\|^{2}-v^2),\label{bogo1}\\
&&D_{i}\phi\mp{i}\sqrt{\gamma}\epsilon_{ij}\gamma^{jk}D_{k}\phi=0.
\label{bogo2}
\end{eqnarray}
The complex Higgs field $\phi$ can be regarded as the complex
representation of a two-dimensional vector field
$\vec{\phi}=(\phi^{1}, \phi^{2})$ over the base space, it is
actually a section of a complex line bundle on the base manifold.
Substituting $\phi=\phi^{1}+i\phi^{2}$ and
$D_{i}=\partial_{i}-ieA_{i}$ into Eq. (\ref{bogo2}) and splitting
the real part form the imaginary part, we obtain two equations
\begin{eqnarray}
\partial_{i}\phi^{j}=-eA_{i}\epsilon_{jk}\phi^{k}
\mp\sqrt{\gamma}\epsilon_{il}\gamma^{lm}(\epsilon_{jk}\partial_{m}\phi^{k}
-eA_{m}\phi^{j}),\label{split}
%&&\partial_{i}\phi^{1}=-eA_{i}\phi^{2}
%\mp\sqrt{\gamma}\epsilon_{ij}\gamma^{jk}(+\partial_{k}\phi^{2}
%-eA_{k}\phi^{1}),\label{split1}\\
%&&\partial_{i}\phi^{2}=+eA_{i}\phi^{1}
%\mp\sqrt{\gamma}\epsilon_{ij}\gamma^{jk}(-\partial_{k}\phi^{1}
%-eA_{k}\phi^{2}).\label{split2}
\end{eqnarray}
From Eq. (\ref{split}), by calculating
$\partial_{i}\phi^{*}\phi-\partial_{i}\phi\phi^{*}$, we can obtain
the expression of the gauge potential
\begin{eqnarray}
eA_{i}=-\frac{1}{2i\|\phi\|^{2}}(\partial_{i}\phi^{*}\phi-\partial_{i}\phi\phi^{*})
\mp \sqrt{\gamma}\epsilon_{ij}\gamma^{jk}\partial_{k}\ln\|\phi\|.
\label{eAi1}
\end{eqnarray}
If we define the unit vector
\begin{equation}\label{na}
n^{a}=\frac{\phi^{a}}{\|\phi\|}, \;\;\;\;(a,b=1,2)
\end{equation}
and note the identity
\begin{equation}\label{identity}
\epsilon_{ab}n^{a}\partial_{i}n^{b}=\frac{1}{2i\|\phi\|^{2}}
(\partial_{i}\phi^{*}\phi-\partial_{i}\phi\phi^{*}),
\end{equation}
Eq. (\ref{eAi1}) further simplifies to:
\begin{equation}
eA_{i}=-\epsilon_{ab}n^{a}\partial_{i}n^{b}
\mp\sqrt{\gamma}\epsilon_{ij}\gamma^{jk}\partial_{k}\ln\|\phi\|.\label{eAi2}
\end{equation}
In curved space, the magnetic field is defined by
$B=-\frac{1}{\sqrt{\gamma}}\epsilon^{ij}\partial_{i}{A}_{j}$,
according to Eq. (\ref{eAi2}), we have
\begin{equation}\label{B=delta+ln}
e\sqrt{\gamma}B=\epsilon^{i
j}\epsilon_{ab}\partial_{i}n^{a}\partial_{j}n^{b}
\pm\epsilon^{ij}\epsilon_{jk}\partial_{i}(\sqrt{\gamma}\gamma^{kl}\partial_{l}\ln\|\phi\|).
\end{equation}
So the first self-duality equation (\ref{bogo1}) can be
generalized to
\begin{equation}
\mp e^2 \sqrt{\gamma} (\|\phi\|^{2}-v^2) =
\epsilon^{ij}\epsilon_{ab}\partial_{i}n^{a}\partial_{j}n^{b}
\pm\epsilon^{ij}\epsilon_{jk}\partial_{i}(\sqrt{\gamma}\gamma^{kl}
\partial_{l}\ln\|\phi\|).\label{nonlinear01}
\end{equation}
According to Duan's $\phi$-mapping topological current theory
\cite{DuanKT1976}, it is easy to see that the first term on the
RHS of Eq. (\ref{nonlinear01}) bears a topological origin, and the
topological term just describes the non-trivial distribution of
$\vec{n}$ \cite{'tHooft}. Noticing
$\partial_{i}n^{a}={\partial_{i}\phi^{a}}/{\parallel\phi\parallel}
+\phi^{a}\partial_{i}({1}/{\parallel\phi\parallel})$ and the Green
function relation in $\phi$-space:
$\partial_{a}\partial_{a}ln(\|\phi\|)=2\pi\delta^{2}(\vec{\phi}\;),\;
(\partial_{a}={\frac{\partial}{\partial\phi^{a}}})$, it can be
proved that \cite{DuanNPB1998514}
\begin{equation}
\epsilon^{i j}\epsilon_{ab}\partial _{i}n^{a}\partial_{j}n^{b}
=2\pi\delta^{2}(\vec{\phi}\;)J(\frac{\phi}{y})
=2\pi\sum_{k=1}^NW_k\delta (\vec{y}-\vec{y}_{k}), \label{BT=del}
\end{equation}
where $J(\phi/y)$ is the Jacobian and $W_k=\beta _k\eta _k$ is the
winding number around the $k$-th vortex, the positive integer
$\beta_k$ is the Hopf index and $\eta _k=\pm 1$ is the Brouwer
degree, $\vec{y}_{k}$ are the coordinates of the $k$-th vortex. So
the first Bogomol'nyi self-duality equation (\ref{bogo1}) should
be
\begin{eqnarray}
\mp e^2 \sqrt{\gamma} (\|\phi\|^{2}-v^2)
=2\pi\sum_{k=1}^NW_k\delta (\vec{y}-\vec{y}_{k}) ~\mp~
\epsilon^{ij}\epsilon_{jk} \partial_{i} (\sqrt{\gamma} \gamma^{kl}
\partial_{l} \ln\|\phi\|).\label{generaleq}
\end{eqnarray}
Obviously the first term on the RHS of Eq. (\ref{generaleq})
describes the topological self-dual vortex.

Now let us discuss the case of flat space for the self-duality
equation (\ref{generaleq}). In this special case,
$\gamma_{ij}=\delta_{ij}$ and Eq. (\ref{generaleq}) reads as
\begin{equation}
\mp e^2(\|\phi\|^{2}-v^2) %
=2\pi\sum_{k=1}^NW_k\delta (\vec{y}-\vec{y}_{k}) %
~\mp~\partial_{i}\partial_{i}\ln\|\phi\|.\label{flateq}
\end{equation}
While the corresponding conventional self-duality equation is
\cite{Dunne1998}
\begin{equation}
e^2(\|\phi\|^{2}-v^2)
=\partial_{i}\partial_{i}\ln\|\phi\|.\label{conventionalEq}
\end{equation}
Comparing our equation (\ref{flateq}) with Eq.
(\ref{conventionalEq}), one can see that the topological term $ 2
\pi {\sum_{k=1}^N} {W_k} \delta (\vec{y} - \vec{y}_{k})$, which
describes the topological self-dual vortex, is missed in the
conventional equation. Obviously, only when the field $\phi\neq0$,
the topological term vanishes and the conventional equation is
correct. So, the exact self-duality equation should be Eq.
(\ref{flateq}) for flat space and Eq. (\ref{generaleq}) for curved
one. As for conventional self-dual nonlinear equation
(\ref{conventionalEq}), a great deal of work has been done by many
physicists on it, and a vortex-like solution was given by Jaffe
\cite{jaffe}. But no exact solutions are known.

In the following sections, we first give a brief review of a
string-like defect solution to Einstein's equations with sources,
then we study fermionic zero modes coupled with the vortex
background and the localization of fermions on the string-like
defect with an exponentially decreasing warp factor.

\section{Review of a string-like defect}

Let us consider Einstein's equations with a bulk cosmological
constant $\Lambda$ and an energy-momentum tensor $T_{MN}$ in
general six dimensions:
\begin{eqnarray}
R_{MN} - \frac{1}{2} g_{MN} R = - \Lambda g_{MN}  + \kappa_6^2
T_{MN}, \label{EinsteinEq}
\end{eqnarray}
where $\kappa_6$ denotes the 6-dimensional gravitational constant
with a relation $\kappa_6^2 = 8 \pi G_N = {8 \pi}/ {M_*^{4}}$,
$G_N$ and $M_*$ being the 6-dimensional Newton constant and the
6-dimensional Planck mass scale, respectively, the energy-momentum
tensor is defined as
\begin{eqnarray}
T_{MN} = - \frac{2}{\sqrt{-g}} \frac{\delta}{\delta g^{MN}} \int
d^6 x \sqrt{-g} L_m.
\end{eqnarray}
%% Throughout this article we follow the standard conventions and
%% notations of the textbook of Misner, Thorne and Wheeler
%% \cite{Misner}.

We shall consider the most general metric ansatz for a warped
brane embedded in six dimensions obeying four dimensional Poincare
invariance
%We shall consider 6-dimensional manifolds with the geometry
\begin{eqnarray}
 ds^2 %&=& g_{MN} dx^M dx^N  \nn\\
      &=& e^{-A(r)} \hat{g}_{\mu\nu}(x) dx^\mu dx^\nu
       + dr^2
       +R_0^2 ~ e^{-B(r)} d\theta^{2}, \label{ds2}
\end{eqnarray}
where the radial coordinate $r$ is infinitely extended
($0<r<\infty$) and the compact coordinate $\theta$ ranges from $0
\leq \theta \leq 2\pi$, $R_0$ is an additional parameter
characterizing the extra compact direction. Moreover, we shall
adopt the ansatz for the energy-momentum tensor respecting the
spherical symmetry:
\begin{eqnarray}
  T^\mu_\nu = \delta^\mu_\nu t_{0}(r),  ~~
  T^r_r = t_{r}(r), ~~
  T^\theta_\theta = t_{\theta}(r),  \label{TMN}
\end{eqnarray}
where $t_i(i=0, r, \theta)$ are functions of only the radial
coordinate $r$.

Under these ansatzs, Einstein's equations (\ref{EinsteinEq}) and
the conservation law for energy-momentum tensor $\nabla^M T_{MN} =
0$ reduce to
\begin{eqnarray}
 e^A \hat{R} - 3 (A')^2
      -2 A' B'
   - 2\Lambda + 2 \kappa_{6}^2 t_{r} = 0, \\
 e^A \hat{R} + 4A''
     - 5(A')^2  - \frac{1}{2} (B')^2
     - 2\Lambda + 2 \kappa_{6}^2 t_{\theta} = 0, \\
e^A \hat{R} + 2 B'' + 6 A'' - 10(A')^2 - (B')^2
     - 4\Lambda + 4 \kappa_{6}^2 t_{0} = 0, \\
 t'_r = 2 A' (t_{r} - t_{0})
  + \frac{1}{2} B' (t_{r} - t_{\theta}),
\end{eqnarray}
where $\hat{R}$ are the scalar curvatures associated with the
metric $\hat{g}_{\mu\nu}$, and the prime denotes the derivative
with respect to $r$. Here we define the cosmological constant
$\hat\Lambda$ on the 3-brane by the equation
\begin{eqnarray}
 \hat{R}_{\mu\nu} - \frac{1}{2} \hat{g}_{\mu\nu} \hat{R}
   = - \hat\Lambda \hat{g}_{\mu\nu}.
\end{eqnarray}

It is now known that there are many interesting solutions to these
equations (see, for instance, \cite{Vilenkin}). Here, we shall
consider the brane solutions with a warp factor
\begin{eqnarray}
 A(r) = c r, \label{12}
\end{eqnarray}
where $c$ is a constant. A specific solution occurs when we have
the spontaneous symmetry breakdown $t_{r} = -t_{\theta}$
\cite{Vilenkin}:
\begin{eqnarray}
ds^2 = e^{-cr} \hat{g}_{\mu\nu} dx^\mu dx^\nu + dr^2 + R_0^2 ~
e^{-c_1 r} d\theta^2, \label{StringLikeSolution}
\end{eqnarray}
where
\begin{eqnarray}
 c^2 &=& \frac{2}{5}(\kappa_{6}^2 t_{\theta} -\Lambda) > 0, \\
 c_1 &=& c - \frac{2}{c} \kappa_{6}^2 t_{\theta}, \\
 \hat{R} &=& 4 \hat{\Lambda} = 0. \label{c_c1_hatR}
\end{eqnarray}
This special solution would be utilized to analyze localization of
fermionic fields on a string-like defect.

\section{Localization of fermions}

In this section, we have the physical setup in mind such that
`local cosmic string' sits at the origin $r=0$ and then ask the
question of whether various bulk fermions with spin 1/2 and 3/2
can be localized on the brane with the exponentially decreasing
warp factor by means of the gravitational interaction and vortex
backgrounds. Of course, we have implicitly assumed that various
bulk fields considered below make little contribution to the bulk
energy so that the solution (\ref{StringLikeSolution}) remains
valid even in the presence of bulk fields.

\subsection{Spin 1/2 fermionic field}\label{A}

In this subsection we study localization of a spin 1/2 fermionic
field in gravity (\ref{StringLikeSolution}) and vortex
backgrounds. It will be shown that provided that if the vortex
background satisfies certain condition, there is a localized zero
mode on the string-like defect.

Let us consider the Dirac action of a massless spin 1/2 fermion
coupled to gravity and vortex backgrounds:
\begin{eqnarray}
 S_m = \int d^D x \sqrt{-g}
       i \bar{\Psi} \Gamma^M D_M \Psi
 , \label{DiracAction}
\end{eqnarray}
from which the equation of motion is given by
\begin{eqnarray}
\Gamma^M  ( \partial_M + \omega_M  -ie A_M ) \Psi=0,
\end{eqnarray}
where $\omega_M= \frac{1}{4} \omega_M^{\bar{M} \bar{N}}
\Gamma_{\bar{M}} \Gamma_{\bar{N}}$ is the spin connection with
$\bar{M}, \bar{N}, \cdots$ denoting the local Lorentz indices,
$\Gamma^M$ and $\Gamma^{\bar{M}}$ are the curved gamma matrices
and the flat ones, respectively. From the formula $\Gamma^M = e^M
_{\bar{M}} \Gamma^{\bar{M}}$ with $e_M ^{\bar{M}}$ being the
vielbein, we have the relations:
\begin{eqnarray}
 \Gamma^\mu = \mathrm{e}^{\frac{1}{2}cr}
              \hat{e}^{\mu}_{\bar{\mu}} \Gamma^{\bar{\mu}},~~
%            = \mathrm{e}^{\frac{1}{2}cr}\hat{\Gamma}^{\mu},~~
 \Gamma^r   = \delta^r_{\bar{r}} \Gamma^{\bar{r}}, ~~
 \Gamma^\theta =R_0^{-1} \mathrm{e}^{\frac{1}{2}c_{1}r} \delta^\theta_{\bar{\theta}}
             \Gamma^{\bar{\theta}}. \label{GammaMatrices}
\end{eqnarray}
The spin connection $\omega_M^{\bar{M} \bar{N}}$ in the covariant
derivative $D_M \Psi$ is defined as
\begin{eqnarray}
 \omega_M ^{\bar{M} \bar{N}}
 &=& \frac{1}{2} e^{N \bar{M}}(\partial_M e_N ^{\bar{N}}
                      - \partial_N e_M ^{\bar{N}}) \nn\\
 &-& \frac{1}{2} e^{N \bar{N}}(\partial_M e_N ^{\bar{M}}
                      - \partial_N e_M ^{\bar{M}}) \nn\\
 &-& \frac{1}{2} e^{P \bar{M}} e^{Q \bar{N}} (\partial_P e_{Q
{\bar{R}}} - \partial_Q e_{P {\bar{R}}}) e^{\bar{R}} _M.
\label{55}
\end{eqnarray}
So the non-vanishing components of $\omega_M$ are
\begin{eqnarray}
  \omega_\mu = \frac{1}{4}c \Gamma_r\Gamma_\mu,~~~
  \omega_\theta = \frac{1}{4} c_1 \Gamma_r
            \Gamma_\theta.
\end{eqnarray}

In what follows, to illustrate how the vortex background affects
the fermionic zero modes, we first discuss the simple case that
the Higgs field $\phi$ is only relative to $r$, and then solve the
general Dirac equation for the vacuum Higgs field solution
$\|\phi\|=v$.

Case I: $\phi=\phi(r)=\phi^1 (r)+i\phi^2 (r)$.

In this case, Eq. (\ref{eAi2}) reduces to:
\begin{eqnarray}
eA_{r}&=&-\epsilon_{ab}n^{a}
\partial_{r}n^{b},\label{A1CaseI}\\
eA_{\theta}&=&\pm R_{0}^{-1} \mathrm{e}^{\frac{1}{2}c_{1}r}
\partial_{r}\ln\|\phi\|.\label{A2CaseI}
\end{eqnarray}
The Dirac equation then becomes
\begin{equation}
 \left\{\mathrm{e}^{\frac{1}{2}cr} \hat{e}^{\mu}_{\bar{\mu}}
         \Gamma^{\bar{\mu}} \hat{D}_{\mu}
        +\Gamma^r \left( \partial_r - c - \frac{1}{4}c_{1}
                + i\epsilon_{ab}n^{a} \partial_{r}n^{b}
                \mp iR_{0}\Gamma^{r}\Gamma^{\theta}
                R_{0}^{-1} \mathrm{e}^{\frac{1}{2}c_{1}r}
                \partial_{r}\ln\|\phi\|\right)
        +\Gamma^{\theta}\partial_\theta
         \right \} \Psi=0, \label{DiracEq2}
\end{equation}
where $\hat{e}^{\mu}_{\bar{\mu}} \Gamma^{\bar{\mu}}
\hat{D}_{\mu}=\hat{e}^{\mu}_{\bar{\mu}} \Gamma^{\bar{\mu}}
({\partial_\mu-ieA_{\mu}})$ is the Dirac operator on the
4-dimensional braneworld in the background of the gauge field
$A_\mu$. We are now ready to study the above Dirac equation for
6-dimensional fluctuations, and write it in terms of 4-dimensional
effective fields. Since $\Psi$ is a 6-dimensional Weyl spinor we
can represent it by \cite{Parameswaran0608074}
\begin{equation}
 \Psi=\left(%
\begin{array}{c}
  \Psi^{(4)} \\
  0 \\
\end{array}%
\right),
\end{equation}
where $\Psi^{(4)}$ is a 4-dimensional Dirac spinor. Our choice for
the 6-dimensional constant gamma matrices $\Gamma^{\bar{M}}$,
$M=0,1,2,3,\bar{r},\bar{\theta}$ are
\begin{equation}
\Gamma^{\bar{\mu}}=
\left(%
\begin{array}{cc}
  0 & \gamma^{\bar{\mu}} \\
  \gamma^{\bar{\mu}} & 0 \\
\end{array}%
\right),~~
\Gamma^{\bar{r}}=
\left(%
\begin{array}{cc}
  0 & \gamma^{5} \\
  \gamma^{5} & 0 \\
\end{array}%
\right),~~
\Gamma^{\bar{\theta}}=
\left(%
\begin{array}{cc}
  0 & -i \\
  i & 0 \\
\end{array}%
\right), \label{Gamma}
\end{equation}
where the $\gamma^{\bar{\mu}}$ are the 4-dimensional constant
gamma matrices and $\gamma^5$ the 4-dimensional chirality matrix.
Imposing the chirality condition $\gamma^5 \Psi^{(4)} = +
\Psi^{(4)}$, the Dirac equation (\ref{DiracEq2}) can be written as
\begin{equation}
 \left\{ \mathrm{e}^{\frac{1}{2}cr} \hat{e}^{\mu}_{\bar{\mu}}
         \gamma^{\bar{\mu}} \hat{D}_{\mu}
        +\left( \partial_r - c - \frac{1}{4}c_{1}
                +i\epsilon_{ab}n^{a} \partial_{r}n^{b}
                \pm R_0^{-2}\mathrm{e}^{c_{1}r}\partial_{r}\ln\|\phi\|
         \right)
        +i R_0^{-1} \mathrm{e}^{\frac{1}{2}c_{1}r}
        \partial_\theta  \right \} \Psi^{(4)}=0. \label{DiracEq3}
\end{equation}

Now, from the equation of motion (\ref{DiracEq3}), we will search
for the solutions of the form
\begin{equation}
 \Psi^{(4)}(x,r,\theta) = \psi(x) \alpha(r) \sum \mathrm{e}^{il \theta},
\end{equation}
where $\psi(x)$ satisfies the massless 4-dimensional Dirac
equation $\hat{e}^{\mu}_{\bar{\mu}}\gamma^{\bar{\mu}}
\hat{D}_{\mu} \psi = 0$. For $s$-wave solution, Eq.
(\ref{DiracEq3}) is reduced to
\begin{eqnarray}
 \left( \partial_r - c - \frac{1}{4}c_{1}
        +i\epsilon_{ab}n^{a}\partial_{r}n^{b}
        \pm R_0^{-2}\mathrm{e}^{c_{1}r}\partial_{r}\ln\|\phi\|
 \right)
 \alpha(r) = 0.
\end{eqnarray}
The solution of this equation is given by
\begin{eqnarray}
 \alpha(r) \varpropto  {\exp}
     \left\{ cr + {1\over 4}c_{1}r
             - i\int^r dr \epsilon_{ab}n^{a}\partial_{r}n^{b}
            \mp R_0^{-2}\int^r dr \mathrm{e}^{c_{1}r} \partial_{r}\ln\|\phi\|
     \right\}. \label{ZeroMode}
\end{eqnarray}
So the fermionic zero mode reads
\begin{eqnarray}
\Psi \varpropto \left(%
\begin{array}{c}
  \psi \\
  0 \\
\end{array}%
\right)
 {\exp}\left\{ cr + {1\over 4}c_{1}r
               - i\int^r dr \epsilon_{ab}n^{a}\partial_{r}n^{b}
               \mp R_0^{-2}\int^r dr \mathrm{e}^{c_{1}r}
                   \partial_{r}\ln\|\phi\|
       \right\}. \label{ZeroMode1}
\end{eqnarray}

Now we wish to show that this zero mode is localized on the defect
sitting around the origin $r=0$ under certain conditions. The
condition for having localized 4-dimensional fermionic field is
that $\alpha(r)$ is normalizable. It is of importance to notice
that normalizability of the ground state wave function is
equivalent to the condition that the ``coupling" constant is
nonvanishing.

Substituting the zero mode (\ref{ZeroMode1}) into the Dirac action
(\ref{DiracAction}), the effective Lagrangian for $\psi$ then
becomes
\begin{eqnarray}
 \mathcal{L}_{eff}^{(0)}
  &=& \int drd\theta \sqrt{-g}
      \bar{\Psi} i \Gamma^M D_M \Psi \nn\\
  &=&  I_{1/2} ~ \sqrt{-\hat{g}}~\bar{\psi}
      i \hat{e}^{\mu}_{\bar{\mu}}\gamma^{\bar{\mu}} \hat{D}_{\mu} \psi, \label{Leff1}
\end{eqnarray}
where
\begin{eqnarray}
 I_{1/2} &\varpropto&   \int_0^{\infty} dr
     \exp\left(\frac{1}{2} c r
               \mp 2 R_0^{-2}\int^r dr \mathrm{e}^{c_{1}r}
                   \partial_{r}\ln\|\phi\|
         \right). \label{I12CaseI}
\end{eqnarray}
In order to localize spin 1/2 fermion in this framework, the
integral (\ref{I12CaseI}) should be finite.  From Eq.
(\ref{I12CaseI}), one can see that whatever the form of $A_r(r)$
is, the effective Lagrangian for $\psi(x)$ has the same form. If
the vortex background vanishes, this integral is obviously
divergent for $c>0$ while it is finite for $c<0$. If the vortex
background does not vanish, the requirement that the integral
(\ref{I12CaseI}) should be finite for $c>0$ is easily satisfied.
For example, a simple choice is $\|\phi\| = \mathrm{e}^{{\pm} r}$
and $c_1>0$. These fermionic zero modes are generically
normalizable on the brane in the self-dual vortex background  if
the integral $I_{{1}/{2}}$ does not diverge.

Case II: the vacuum solution $\|\phi\|^2=v^2$.

For the vacuum solution, $\|\phi\|^2=v^2$, i.e.
$\phi=ve^{i\theta}$, we have $ n^{r}=\cos\theta$,
$n^{\theta}=\sin\theta$, and $eA_{r}=0$, $eA_{\theta}=-1$. The
Dirac equation then becomes
\begin{equation}
 \left\{ \mathrm{e}^{\frac{1}{2}cr} \hat{e}^{\mu}_{\bar{\mu}}
         \Gamma^{\bar{\mu}} \hat{D}_{\mu}
        +\Gamma^r \left( \partial_r - c
                         - \frac{1}{4}c_{1}\right)
        +\Gamma^{\theta}(\partial_\theta+i) \right \} \Psi=0. \label{DiracEq2'}
\end{equation}
Repeating the deduction as the above case and imposing the
chirality condition $\gamma^5 \Psi^{(4)} = -\Psi^{(4)}$, one can
get the fermionic zero modes
\begin{eqnarray}
\Psi \varpropto \left(%
\begin{array}{c}
  \psi \\
  0 \\
\end{array}%
\right)
 {\exp}\left\{ cr + {1\over 4}c_{1}r
               - R_0^{-1} \int^r dr ~ \mathrm{e}^{\frac{1}{2}c_{1}r}
       \right\}. \label{ZeroMode2}
\end{eqnarray}
The effective Lagrangian for $\psi(x)$ then becomes
\begin{eqnarray}
 \mathcal{L}_{eff}^{(0)}
  &=& \int drd\theta \sqrt{-g}
      \bar{\Psi}_0 i \Gamma^M D_M \Psi_0 \nn\\
  &=&  I_{1/2} ~ \sqrt{-\hat{g}}~\bar{\psi}
      i \gamma^{\mu} \hat{D}_{\mu} \psi, \label{Leff1}
\end{eqnarray}
where
\begin{eqnarray}
 I_{1/2} &\varpropto&   \int_0^{\infty} dr
     \exp\left(\frac{1}{2} c r
               -2 R_0^{-1} \int^r dr ~ \mathrm{e}^{\frac{1}{2}c_{1}r}
     \right). \label{I12CaseII}
\end{eqnarray}
In order to localize spin 1/2 fermion on a string-like defect with
the exponentially decreasing warp-factor (i.e. $c>0$) in this
framework, the integral (\ref{I12CaseII}) should be finite. It is
easy to see that the condition is $c_{1}>0$, i.e.
\begin{eqnarray}
 \begin{array}{l}
   \Lambda < -4\kappa_{6}^2 t_{\theta}, ~~~\;\;
           \mathrm{for}~~ t_{\theta} > 0 \\
   \Lambda < \kappa_{6}^2 t_{\theta}. ~~~~~~~~~
           \mathrm{for}~~ t_{\theta} < 0 \\
 \end{array}
 \label{conditionCaseII}
\end{eqnarray}
This situation is a little different from the above case.

\subsection{Spin 3/2 fermionic field}

Next we turn to spin 3/2 field, in other words, the gravitino. Let
us start by considering the action of the Rarita-Schwinger
gravitino field:
\begin{eqnarray}
S_m = \int d^D x \sqrt{-g} \bar{\Psi}_M i \Gamma^{[M} \Gamma^N
\Gamma^{R]} D_N \Psi_R, \label{GravitinoAction}
\end{eqnarray}
where the square bracket denotes the anti-symmetrization, and the
covariant derivative is defined with the affine connection
$\Gamma^R_{MN} = e^R_{\bar{M}}(\partial_M e_N^{\bar{M}} +
\omega_M^{\bar{M} \bar{N}} e_{N {\bar{N}}})$ by
\begin{eqnarray}
 D_M \Psi_N = \partial_M \Psi_N - \Gamma^R_{MN} \Psi_R
            + \omega_M \Psi_N + A_M \Psi_N. \label{64}
\end{eqnarray}
From the action (\ref{GravitinoAction}), the equations of motion
for the Rarita-Schwinger gravitino field are given by
\begin{eqnarray}
\Gamma^{[M} \Gamma^N \Gamma^{R]} D_N \Psi_R = 0.
\label{GravitinoEq}
\end{eqnarray}

For simplicity, from now on we limit ourselves to the flat brane
geometry $\hat{g}_{\mu\nu} = \eta_{\mu\nu}$. After taking the
gauge condition $\Psi_\theta = 0$ and $A_\mu=0$, the non-vanishing
components of the covariant derivative are calculated as follows:
\begin{eqnarray}
 D_\mu \Psi_\nu &=& \partial_\mu \Psi_\nu
     - \frac{1}{2} c e^{-cr} \eta_{\mu\nu} \Psi_r
     + \frac{1}{4} c \Gamma_r \Gamma_\mu \Psi_\nu
     -ie A_{\mu} \Psi_{\nu}, \nn\\
 D_\mu \Psi_r &=& \partial_\mu \Psi_r
     + \frac{1}{2} c\Psi_\mu
     + \frac{1}{4} c \Gamma_r\Gamma_\mu \Psi_r
     -ie A_{\mu} \Psi_{r} , \nn\\
 D_r \Psi_\mu &=& \partial_r \Psi_\mu
     + \frac{1}{2} c\Psi_\mu
     -ie A_{r} \Psi_{\mu}, \nn\\
 D_r \Psi_r &=& \partial_r \Psi_r -ie A_{r} \Psi_r, \label{derivative} \\
 D_\theta \Psi_\mu &=& \partial_\theta \Psi_\mu
     + \frac{1}{4} c_1 \Gamma_r \Gamma_\theta \Psi_\mu
     -ie A_{\theta} \Psi_{\mu}, \nn\\
 D_\theta \Psi_r &=& \partial_\theta \Psi_r
     + \frac{1}{4} c_1 \Gamma_r \Gamma_\theta \Psi_r
     -ie A_{\theta} \Psi_{r}, \nn\\
 D_\theta \Psi_\theta &=& -\frac{1}{2} c_1 R_0 e^{-c_1 r} \Psi_r.
 \nn
\end{eqnarray}
Again we  represent $\Psi_M$ as the following form
\begin{equation}
 \Psi_{M}=\left(%
\begin{array}{c}
  \Psi_{M}^{(4)} \\
  0 \\
\end{array}%
\right),\label{Psimu}
\end{equation}
where $\Psi_{M}^{(4)}$ is the 4D Rarita-Schwinger gravitino field.

Imposing the chirality condition $\gamma^5 \Psi_{\mu}^{(4)} = +
\Psi_{\mu}^{(4)}$, and substituting Eqs. (\ref{derivative}) and
(\ref{Psimu}) into the equations of motion (\ref{GravitinoEq}), we
will look for the solutions of the form
\begin{eqnarray}
 \Psi_\mu^{(4)}(x,r,\theta) &=& \psi_\mu(x) u(r) \sum \mathrm{e}^{il
 \theta},\\
 \Psi_r^{(4)}(x,r,\theta) &=& \psi_r(x) u(r) \sum \mathrm{e}^{il \theta},
\end{eqnarray}
where $\psi_\mu(x)$ satisfies the following 4-dimensional
equations $\gamma^\mu \psi_\mu = \partial^\mu \psi_\mu =
\gamma^{[\mu} \gamma^\nu \gamma^{\rho]} (\partial_\nu -ie
A_\nu)\psi_\rho = 0$. Then the equations of motion
(\ref{GravitinoEq}) reduce to
\begin{eqnarray}
 \left(\partial_r -  \frac{1}{2} c - \frac{1}{4} c_1 -ie A_r(r)
       + e R_0^{-1} \mathrm{e}^{\frac{1}{2}c_{1}r} A_{\theta}(r)
 \right) u(r) = 0,
\end{eqnarray}
from which $u(r)$ is easily solved to be
\begin{eqnarray}
 u(r) \varpropto ~ {\exp}
      \left\{ \frac{1}{2} cr + {1\over 4}c_{1}r
        + ie \int^r dr A_r(r)
        - e R_0^{-1} \int^r dr ~ \mathrm{e}^{\frac{1}{2}c_{1}r} A_{\theta}(r)
      \right\}. \label{ZeroMode2}
\end{eqnarray}
In the above we have considered the $s$-wave solution and $\psi_r
= 0$.

Let us substitute the zero mode (\ref{ZeroMode2}) into the
Rarita-Schwinger action (\ref{GravitinoAction}). It turns out that
the effective Lagrangian becomes
\begin{eqnarray}
 \mathcal{L}_{eff}
   &=& \int drd\theta \sqrt{-g} \bar{\Psi}_M i \Gamma^{[M}
       \Gamma^N \Gamma^{R]} D_N \Psi_R \nn\\
   &=& I_{3/2} ~\bar{\psi}_\mu i \gamma^{[\mu} \gamma^\nu
       \gamma^{\rho]} (\partial_\nu -ie A_\nu)\psi_\rho, \label{L32}
\end{eqnarray}
where the integral $I_{3/2}$ is defined as
\begin{eqnarray}
 I_{3/2} &\varpropto&   \int_0^{\infty} dr \exp
    \left(\frac{1}{2} c r  - 2e R_0^{-1} \int^r dr ~
       \mathrm{e}^{\frac{1}{2}c_{1}r} A_{\theta}(r)
     \right). \label{I32}
\end{eqnarray}

As in the above subsection, to illustrate how the vortex
background affects the fermionic zero modes, we first discuss the
simple case that the Higgs field $\phi$ is only relative to $r$,
and then solve the general Dirac equation for the vacuum Higgs
field solution $\|\phi\|=v$.

Case I: $\phi=\phi(r)=\phi^1 (r)+i\phi^2 (r)$.

In this case, Eq.(\ref{eAi2}) reduces to Eqs. (\ref{A1CaseI}) and
(\ref{A2CaseI}), and the integral $I_{3/2}$ can be expressed as
\begin{eqnarray}
 I_{3/2} \varpropto  \int_0^{\infty} dr \exp
    \left(\frac{1}{2} c r
          \mp 2 R_{0}^{-2} \int^r dr e^{c_1 r}\partial_{r}\ln \|\phi\|
     \right). \label{I32CaseI}
\end{eqnarray}
It is easy to see that this expression is equivalent to $I_{1/2}$
in (\ref{I12CaseI}) up to an overall constant factor so we
encounter the same result as in the corresponding case for spin
1/2 field. So the zero modes for spin 3/2 field are generically
normalizable on the brane in the self-dual vortex background if
the integral $I_{3/2}$ (\ref{I32CaseI}) does not diverge.

Case II: the vacuum solution $\|\phi\|^2=v^2$.

In this case, $eA_{r}=0$, $eA_{\theta}=-1$. Again, changing the
chirality condition to $\gamma^5 \Psi_{\mu}^{(4)} =
-\Psi_{\mu}^{(4)}$, the integral $I_{3/2}$ takes the form
\begin{eqnarray}
 I_{3/2} &\varpropto&   \int_0^{\infty} dr
     \exp\left(\frac{1}{2} c r
               -2 R_0^{-1} \int^r dr ~ \mathrm{e}^{\frac{1}{2}c_{1}r}
     \right), \label{I32CaseII}
\end{eqnarray}
which is equivalent to $I_{1/2}$ in (\ref{I12CaseII}) up to an
overall constant factor so it is also finite for $c>0$ and
$c_{1}>0$.

\section{Discussions}
Using the generalized Abelian Higgs model and $\phi$-mapping
theory, we investigate the self-dual vortex on an extra
two-dimensional curved Riemann surface, and obtain the inner
topological structure of the self-dual vortex. Under the gravity
and vortex backgrounds, we have investigated the possibility of
localizing the spin 1/2 and 3/2 fermionic fields on a brane with
the exponentially decreasing warp factor. We first give a brief
review of a string-like defect solution to Einstein's equations
with sources, then check localization of fermionic fields on such
a string-like defect with the background of self-dual vortex from
the viewpoint of field theory. It has been found that the vortex
background affects the fermionic zero modes, and that spin 1/2 and
3/2 fields can be localized on a defect with the exponentially
decreasing warp factor if self-dual vortex and gravitational
backgrounds are considered.

%For the case that the Higgs field $\phi$ is only relative to $r$,

%For the vacuum solution $\|\phi\|^2=v^2$, the condition is $c>0$
%and $c_{1}>0$.

%These results for localization of various spin fields coincide
%with the corresponding ones \cite{{Bajc}} in the Randall-Sundrum
%model \cite{RS} and many brane models \cite{Oda,Oda2000}. By
%including the self-dual vortex background, it is  remarkable that
%there are more meaningful results owing to the existence of the
%nontrivial exponential factor $c_{1}$ in the angular part of the
%metric.

\section*{Acknowledgement}
This work was supported by the National Natural Science Foundation
of the People's Republic of China and the Fundamental Research
Fund for Physics and Mathematic of Lanzhou University.

\end{document}